\documentclass[grl]{agu2001}
 \usepackage{graphicx}
\authorrunninghead{STARLAB PREPRINT 30052004-01}
\titlerunninghead{Accurate GNSS-R altimetry from aircraft}
\authoraddr{G. Ruffini, Research Department, Starlab, C. de l'Observatori, s/n, 08035 Barcelona, Spain.
(giulio.ruffini@starlab.es)}
\begin{document}

%% ------------------------------------------------------------------------ %%
%
%  TITLE
%
%% ------------------------------------------------------------------------ %%

\title{The Eddy Experiment: accurate GNSS-R ocean altimetry from low altitude aircraft}
%
% e.g., \title{Terrestrial Ring Current:
% Origin, Formation and Decay $\alpha\beta\Gamma\Delta$}

%% ------------------------------------------------------------------------ %%
%
%  AUTHORS AND AFFILIATIONS
%
%% ------------------------------------------------------------------------ %%

% Method 1 (for all journals, except Reviews of Geophysics, which
% should use method 3):
% For three or fewer author/affiliation blocks, use \author{} and \affil{}

%\author{G. Ruffini, F. Soulat, M. Caparrini and O. Germain}
%\affil{ Starlab, C. de l'Observatori Fabra s/n, 08035 Barcelona, Spain, http://starlab.es}

% ---------------
% Method 2 (for all journals, except Reviews of Geophysics, which
% should use method 3): For more than three author/affiliation blocks,
% use \author{\altaffilmark{}} and \altaffiltext{}
% \altaffilmark will produce footnote; matching altaffiltext
% will appear at bottom of page. May use \\ to start a new line.

\authors{G. Ruffini, F. Soulat, M. Caparrini, O. Germain \altaffilmark{1}
 and M. Mart\'{\i}n-Neira \altaffilmark{2}}
% R. Williams, \altaffilmark{3}
% and J. R. McConnell\altaffilmark{4}}

\altaffiltext{1}
{Starlab, C. de l'Observatori Fabra s/n, 08035 Barcelona, Spain, http://starlab.es}
\altaffiltext{2}{ESA/ESTEC, Keplerlaan 1, 2200 Noordwijk, The Netherlands,  http://esa.int}
%
% \altaffiltext{3}{Department of Space Sciences, University of Michigan,
% Ann Arbor, Michigan, USA.}
%
% \altaffiltext{4}{Desert Research Institute, Division of Hydrologic Sciences,
% Reno, Nevada, USA.}

%---------------
% Method 3 (for Reviews of Geophysics only): Reviewauthors is a table with three
% columns. You must supply the ''&'' between each author/affiliation. If you have
% more than three authors, start a new table line with /cr

% e.g.,
% \begin{reviewauthors}
% R. C. Bales\\
% Department of Hydrology and\\ Water Resources\\
% University of Arizona\\
% Tucson, Arizona, USA
% &
% E. Mosley-Thompson\\
% Department of Geography\\
% Ohio State University\\
% Columbus, Ohio, USA
% &
% J. R. McConnell\\
% Desert Research Institute\\
% Reno, Nevada, USA
% \end{reviewauthors}

%% ------------------------------------------------------------------------ %%
%
%  ABSTRACT
%
%% ------------------------------------------------------------------------ %%

% Do NOT include any \begin...\end commands within
% the body of the abstract.

\begin{abstract}
During the Eddy Experiment, two synchronous  GPS receivers were flown at 1~km altitude to collect L1 signals and their reflections from the sea surface for assessment of altimetric precision and accuracy. Wind speed (U10) was around 10~m/s, and SWH up to 2~m. A geophysical parametric waveform model was used for retracking and estimation of  the lapse between the direct and reflected signals with a 1-second precision of 3~m. The lapse was used to estimate the SSH along the  track  using a differential  model.  The RMS error of the 20~km averaged GNSS-R absolute altimetric solution with respect to Jason-1 SSH  and a GPS buoy measurement was of 10~cm, with a 2~cm mean difference. Multipath and retracking  parameter sensitivity due to the low altitude are suspected to have degraded accuracy.  This result provides an important  milestone on the road to a GNSS-R mesoscale altimetry space mission. 
\end{abstract}
%% ------------------------------------------------------------------------ %%
%
%  TEXT
%
%% ------------------------------------------------------------------------ %%

% The body of the article must start with a \begin{article} command,
% and an \end{article} command must follow the references section.
% Otherwise, the text will not print at the appropriate column width.
%

\begin{article}

\section{Introduction}
Several Global Navigation Satellite System (GNSS) constellations and augmentation systems are presently operational or under development, such as the pioneering US Global Positioning System (GPS),  the Russian Global Navigation Satellite System (GLONASS) and the European EGNOS. In the next few years, the European Satellite Navigation System (Galileo) will be deployed, and GPS will be upgraded with more frequencies and civilian codes. By the time Galileo becomes operational, more than 50 GNSS satellites will be emitting very precise L-band spread spectrum signals, and will remain in operation for at least a few decades. Although originally meant for  localization, these signals will no doubt be used within GCOS/GOOS\footnote{Global Climate Observing System/Global Ocean Observing System.} in many ways (e.g., atmospheric sounding). We focus here on the budding field known as GNSS Reflections, which aims at providing instruments and techniques for remote sensing of the ocean surface (in particular, sea surface height and roughness) and the atmosphere over the oceans. 
 
GNSS-R altimetry  (also known as PARIS, Passive Reflectometry Interferometric System \markcite{{\it Mart\'{\i}n-Neira} [1993]}),  the exploitation of reflected Global Navigation Satellite Systems signals for the measurement of Sea Surface Height (SSH), holds the potential to  provide the unprecedented spatio-temporal samplings needed for mesoscale monitoring of ocean circulation. It is at mesoscale where phenomena such as eddies play a fundamental role in the transport of energy and momentum, yet current systems are unable to probe them.

%Many GNSS-R altimetry and scatterometry experiments have been carried out to date, and the list continues to grow thanks to dedicated efforts in the US and Europe. GNSS-R experimental data has now been gathered from  Earth fixed receivers (\cite{caparrini1998,martin-neira2001,treuhaft2001,caparrini2003} among others),  aircraft (\cite{komjathy1998,garrison1998,cardellach2001a,lowe2002,germain2003} among others), stratospheric balloons (\cite{garrison2000,cardellach2003} among others), and from space (\cite{lowe2002b} among others). This work is converging to a unified understanding of the GNSS-R error budget, but so far these experiments have focused on waveform modeling and short term ranging precision. None to date have successfully retrieved a mesoscale altimetric profile as provided by monostatic radar altimeters such as Jason-1.
 
Many GNSS-R experiments have been carried out to date, and the list continues to grow thanks to dedicated efforts in the US and Europe. GNSS-R experimental data has now been gathered from  ground receivers,  aircraft, stratospheric balloons  and from space (see  \markcite{{\it Caparrini et al.} [2003]} for a brief summary of experimental work). While this work is converging into a unified understanding of the GNSS-R error budget,  these experiments have focused on waveform modeling and short term ranging and altimetric precision (e.g., \markcite{{\it Lowe et al.} [2002a,2002b]} and \markcite{{\it Rius et al.} [2002]}). However, none to date have accurately retrieved a mesoscale altimetric profile as provided by monostatic radar altimeters such as Jason-1, which is our aim here.

In this paper we report GNSS-R altimetric results using data from the  Eddy Experiment (09-27-2002). The next section addresses the issue of {\em tracking} the direct and reflected GPS signals, which consist in appropriately placing the delay and Doppler gating windows and in despreading the GPS signals by means of correlation with clean replicas. Tracking produces  incoherently averaged {\em waveforms} (typically with a cadence of 1~second). The extraction of the information needed for the altimetric algorithm from the waveforms is described in the third section. This is the {\em retracking} step, and it yields to the so-called  {\em measured temporal lapse}  (or lapse, for short), in meters, between the direct and reflected signal, i.e., the relative pseudo-range. In the fourth section, the altimetric algorithm  (producing the SSH profile) is described and, finally, results are discussed in the last section.

\begin{figure}[b!]
       \hspace{-0.5cm} \includegraphics[width=8.5cm]{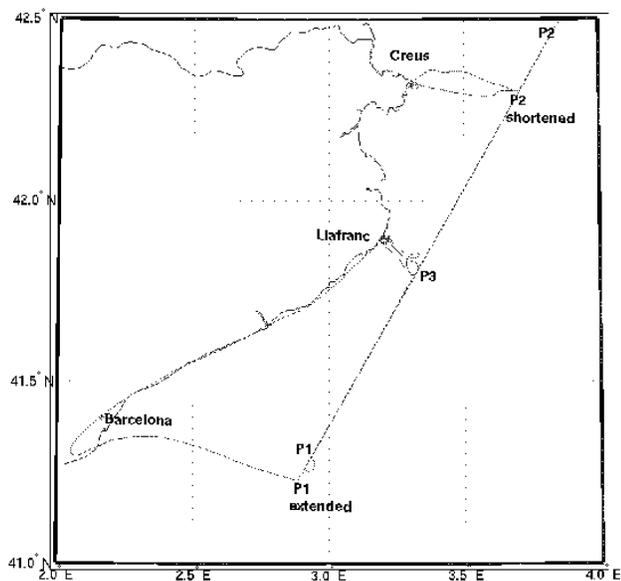} 
    \caption{Total flight trajectory. The data processed corresponds to the ascending track: from P1 to ``P2 shortened''.}
  \label{fig:flight}
\end{figure}

\section{Data collection and pre-processing}
%\subsection{Data set}
The GNSS-R data set was gathered during an  airborne campaign carried out in September~2002. The GPS/INS (Inertial Navigation System) equipped aircraft overflew the Mediterranean Sea, off the coast of Catalonia (Spain), northwards from  the city of Barcelona for about 150~km (Figure~\ref{fig:flight}). As in \markcite{{\it Rius et al.} [2002]},  this area near the Barcelona airport was chosen because it is crossed by a ground track of the Jason-1 altimeter (track number~187).

 %The aircraft  overflew this track during the Jason-1 overpass, for precise comparison. 
%  XXXXXX

The aircraft overflew twice  the Jason-1 ground track, both in the South-North direction and back, gathering about 2 hours of raw GPS-L1~IF data sampled at 20.456~MHz (see \markcite{{\it Soulat} [2003]} for further details). Here we focus on the data processing of the 40 minute long ascending track, i.e., from P1 to P2  in Figure~\ref{fig:flight}, with the Jason-1 overpass coinciding in time roughly during the middle of the track. In addition, a GPS buoy measurement of the SSH on a point along the same track was taken. Given the small tides in the area and the close temporal sampling, tide effects could not introduce  more than 1 cm error in the comparisons of GPS-R and Jason, and no more than 2 cm for GPS-buoy measurements.

%\subsection{Tracking GNSS-R signals}
Altimetry with GNSS-R signals is based on the measurement of the temporal {\em lapse} between the time of arrival of the direct GNSS signal and the time of arrival of the same signal after its reflection from the target surface. Successful tracking of both signals is  the first step for altimetric processing. Under general sea conditions, GPS signals reflected from a rough sea surface cannot be tracked by a standard receiver, because of the signal corruption due to the reflection process. For this reason, a dedicated software receiver has been developed\footnote{The Starlab STARLIGHT processor, also described in \markcite{{\it Caparrini et al.} [2003]}.}.  The processor is composed of two sub-units, one for each signal. The unit which processes the direct signal---the master unit---uses standard algorithms to track the correlation peak of the signal, both in time and frequency. The unit which processes the reflected signal---the slave unit---performs correlations  with  time delay and frequency settings derived from  those of the  master unit. The coherent integration time was set  to 10~milliseconds: it was verified that with this value the ratio of the correlation peak height and the out-of-peak fluctuations  achieved a maximum.  Moreover,  long correlation times  appeared to provide some protection from aircraft multipath by mimicking a higher gain antenna. 

%%%%%%%%%%%%%%%%%%%%%%%%%%%%%%%%%%%%%%%%%%%%%%%%%
\section{Retracking the waveforms}
Once a GNSS-R correlation waveform is obtained from both the direct and reflected signals the lapse  can be estimated. This is not a trivial matter, as the bistatic reflection process deforms severely the reflected signal.
The main challenge for GNSS-R using the GPS Coarse Acquisition (C/A) code is to provide sub-decimetric altimetry using a 300 m equivalent pulse length, something that can only be achieved by modeling and intense averaging with due care of systematic effects. For reference, pulse lengths of modern monostatic altimeters such as Jason-1 are more than two orders of magnitude shorter. 

For these reasons, we estimated the temporal location of each waveform (the {\em delay} parameter)  via a Least Mean Squares model fitting procedure. This is  the so-called {\em retracking} process (as is known in the monostatic altimetric community). The implementation of accurate waveform models (for direct and reflected signals) is fundamental to retracking. Conceptually, a retracking waveform model allows for the transformation of the reflected waveform to an equivalent direct one (or vice-versa), and a  meaningful comparison of direct and reflected waveforms for the lapse estimation.

% \subsection{Waveform model}
The natural  model for retracking  the direct signal waveform is the mean autocorrelation of the GPS C/A code in presence of additive Gaussian noise, which accounts mainly for the receiver noise. As far as the reflected signal is concerned, the model is not so straightforward. In fact, the reflection process induces modifications on the GNSS signals which depend on  sea surface conditions (directional roughness), receiver-emitter-surface kinematics and geometry, and  antenna pattern, all of which have to be taken into account.  In principle, quantities relating to sea surface conditions could be considered as free parameters in the model for the reflected signal waveform, and estimated during the retracking process along with the  delay parameter of the waveform. 

On the basis of the two most quoted models in the literature for bistatic sea-surface scattering  (\markcite{{\it Picardi et al.} [1998]} and \markcite{{\it Zavorotny et al.} [2000]}), we have developed an upgraded waveform model for the reflected signal. This model, as the two aforementioned, is based on the Geometric Optics approximation to the Kirchoff theory---that is to say, with  tangent plane, high frequency and large elevation assumptions, which are reasonable for the quasi-specular sea-surface scattering at L-band (\markcite{{\it Soulat} [2003]}). The main characteristics of this model are  a) the implementation of a fully bistatic geometry (as in \markcite{{\it Zavorotny et al.} [2000]\markcite{{\it Caparrini et al.} [2003]}}, but not in \markcite{{\it Picardi et al.} [1998]}), b) the description of the sea surface through  the L-band Directional Mean Square Slope\footnote{See \markcite{{\it Germain et al.} [2003]} for a discussion on the role of wavelength in the definition of DMSS.} (DMSS$_L$) (as in \markcite{{\it Zavorotny et al.} [2000]}), and c) the use of a fourth parameter, the significant wave height (SWH), to describe the sea surface (as in \markcite{{\it Picardi et al.} [1998]}, but not in \markcite{{\it Zavorotny et al.} [2000]}).
We have checked that the impact of SWH mismodeling in our case is negligible,  since the GPS C/A code equivalent pulse width is about 300 meters. We foresee a higher and non-negligible impact of SWH if the sharper GPS~P-code or future Galileo wide codes are used. 

%\subsection{Inversion scheme}
 Because of the speckle noise affecting the reflected signal waveforms, the retracking fit was not performed on each complex waveform obtained from the 10~ms correlations. Rather, these waveforms were first  incoherently averaged: the average of the magnitude of a hundred 10 ms waveforms was then used to perform the inversion---i.e., 1~second  incoherently averaged real waveforms were generated for retracking. In this way, reflected/direct temporal lapses were produced at a 1 Hz rate. In both cases, the fit of the waveform was performed over three parameters: the lapse,  a scaling factor and the out-of-the-peak correlation mean amplitude.  Retracking of the reflected signal waveform was performed using only the leading edge and a small part of the trailing edge, since the trailing edge is more sensitive to errors in the input parameters (including geophysical parameters and antenna pattern). The geophysical parameters that enter in the model of the reflected signal waveform were not jointly estimated here. These parameters were set to some reasonable {\it a~priori} value obtained from other sources of information (Jason-1 for wind speed, ECMWF for wind direction,  or from theory (for the sea slope PDF isotropy coefficient)---see \markcite{{\it Germain et al., Soulat et al.} [2003]} for more details. 

For convenience, we describe the sea surface state using  wind speed, wind direction and the sea slope PDF isotropy coefficient. Using a spectrum model, these  can be uniquely related to the directional mean square slope parameters as seen in L-band,  DMSS$_L$, with the prior assumption of a mature, wind-driven sea (the sea spectrum in \markcite{{\it Elfouhaily et al.} [1997]} was used in this case). We emphasize that DMSS$_L$ is the actual parameter set needed in the reflection model  under the Geometric Optics approximation. 

Finally, we note that the inclusion of only a small part of the trailing edge makes the retracking algorithm, i.e., the estimation of the three free parameters, more robust with respect to eventual inaccuracies in the value of the fixed parameters. 
%For a sample reflected waveform fit, see Figure~\ref{fig:fit}.

%%%%%%%%%%%%%%%%%%%%%%%%%%%%%%%%%%%%%

\section{Altimetric algorithm}
The output of the retracking process is the time series of measured lapses. The final step  in GNSS-R altimetry is SSH estimation. In order to solve for this quantity, we  used a differential algorithm: a model for the lapse over a reference surface near the local geoid was constructed, and the difference of this reference lapse and the measured one was modeled as  a function of the height over the reference surface.  We note that the aircraft INS data was used to take into account the direct-reflected antenna baseline motion and that we  also included both dry and wet tropospheric delays in the model. Exponential models for tropospheric delays were used with different scale heights and surface values derived from surface  pressure measurements and surface tropospheric delays obtained from ground GPS (see \markcite{{\it Ruffini et al.} [1999]} and references therein) and SSM/I\footnote{Special Sensor Microwave Imager, a passive microwave radiometer flown by the US Defense Meteorological Satellite Program.}.

\begin{figure*}[t!]
   \hspace{0.7cm}     \includegraphics[width=14cm,height=11cm]{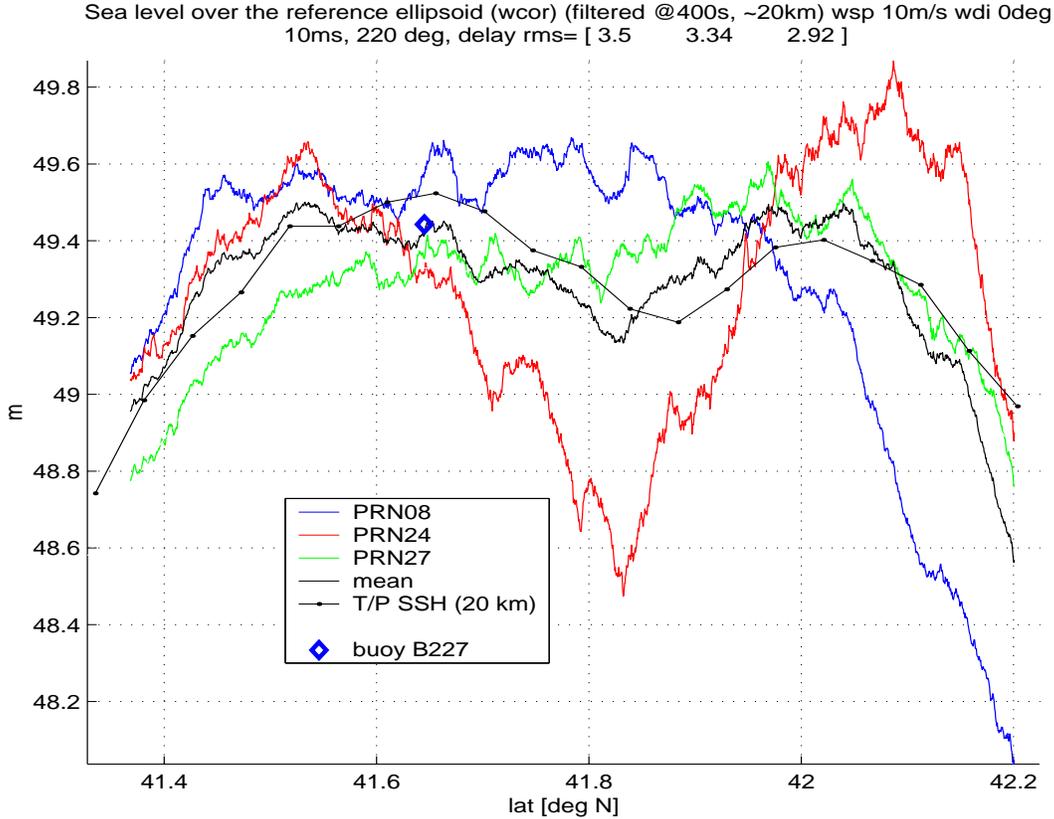} 
    \caption{Final altimetric results. The legend indicates this solution was obtained with a 10 m/s wind speed, 0 degrees direction, 10 ms coherent integration time and nominal antenna pattern. It also indicates the lapse RMS for each PRN, in meters. Individual altimetric results for each satellite are shown in red,  green and blue. The black line represents the average altimetric result of the three GPS satellites boxcar filtered over 20~km---i.e., 400~seconds, the aircraft speed being about~50~m/s---while the dotted, solid black line represents the Jason-1  SSH. The blue diamond is the SSH measurement from the reference buoy.}
  \label{fig:result1}
\end{figure*}

The differential altimetric equation writes
\begin{equation}
\label{eq:diffPARIS}
 \Delta_{DM}\;=\;\Delta_D-\Delta_M\;=\;2\;\delta h\;\sin(\epsilon)+b,
\end{equation}
where $\Delta_D$ is the measured lapse (in meters) as estimated from the data, $\Delta_M$ is the modeled lapse, based on an ellipsoidal model of the Earth, GPS constellation precise ephemeris, aircraft GPS/INS precise kinematic processing, and a tropospheric model,  $\delta h$ is the normal offset between the sea surface  and the (model) ellipsoid surface, $\epsilon$ is the GPS satellite elevation angle at the specular point of reflection, over the ellipsoid, and $b$  is the hardware system bias.
The precision obtained after 1-second of incoherent averaging in the estimation of $\Delta_{DM}$  using this approach 
%is displayed in Table~1.  For each PRN number, the root mean squared error of the 1-second lapse is shown. It is 
was roughly of 3 m, as expected  (given the antenna gain, altitude, etc.) and consistent with earlier experiments.

\section{Results}
The algorithm outlined  in the previous section was implemented to process data from the three better behaved GPS satellites among the four satellites with the highest elevation angle. In fact, one of these four satellites appeared to be severely affected by aircraft-induced multipath (probably due to the landing gear and wing surfaces). A SSH profile was retrieved for each satellite and the average value of the three profiles is shown in Figure~\ref{fig:result1} along with the Jason-1 SSH, Mean Sea Level (MSL) and one SSH value obtained with the control buoy.  This solution was obtained setting a model wind speed of 10 m/s (values provided by Jason-1 along the track vary between 9 and 13~m/s), wind direction of 0 degrees North (values provided by ECMWF vary between  30 and -20~deg North), and  sea slope PDF isotropy coefficient equal to 0.65 (the theoretical value for a mature, wind-driven sea according to \markcite{{\it Elfouhaily et al.} [1997]}).

Inter-satellite altimetric  agreement was moderate, with an average inter-satellite altimetric spread of 10 cm.
% (see Figure~\ref{fig:result12}). 
We believe that discrepancies were caused by multipath from the aircraft surfaces. Hence, the location of the antennas (especially the down-looking one) is an important consideration for future experiments. We also highlight the role of the antenna gain pattern: altimetric solutions appeared to be quite sensitive to rotations of the antenna pattern  (which was not azimuthally symmetric) away from the measured orientation.  The solutions were also quite sensitive to the choice of other input parameters, especially DMSS$_L$. According to our simulations, the impact of all these parameters will decrease rapidly at higher altitudes.

For simplicity, fixed DMSS$_L$ parameters were used along the whole track. It is important to underline that the use of constant values for the geophysical parameters along the whole track (more than 120~km) induces  errors on the final altimetric estimation. Nonetheless, the bias of the final solution with respect to the SSH (the error mean) was 2~cm while the root mean error was 10~cm (Figure~\ref{fig:result1}).

\section{Conclusion}
While the precision of the GNSS-R technique has been analyzed in previous experiments, the novelty here is the demonstration of absolute altimetry.
We have observed that the use of a waveform model for the reflected signal, incorporating parameters to describe the sea surface conditions, is essential for the accuracy of the altimetric solution. The accuracy achieved by the retracking altimetric algorithm used here was of the order of 1 decimeter, with a spatial resolution of 20~km. Future experiments targeting higher chip rate codes (such as the GPS P-code instead of the used C/A-code) should improve these results. Sensitivity analysis has also shown that the altitude of this flight was not optimal for GNSS-R altimetry and made the experiment more prone to aircraft multipath problems.  The angular span of the first-chip zone decreases with altitude---reducing the impact of geophysical parameters, antenna pattern and aircraft multipath on the retracking process of the leading edge---producing a more robust altimetric solution. 
Higher altitude flights are thus now needed, also  to better understand  the GNSS-R space scenario.

We believe the Eddy Experiment is an important milestone on the road to a space mission. We underline that the obtained precision is in line with  earlier experiments and theoretical error budgets (see, e.g., \markcite{{\it Lowe  et al.} [2000a,2000b]}). We note that the same error budgets  have been used to investigate and confirm the strong impact of space-borne GNSS-R altimetric mission data on mesoscale ocean circulation models (\markcite{{\it Letraon et al.} [2003]}). 

Finally, and as discussed in \markcite{{\it Germain et al., Soulat et al.} [2003]}, we would like to emphasize that GNSS-R signals can be profitably used also for surface roughness measurements, and that the two measurements are very synergic.
Altimetric and sea roughness GNSS-R processing can be merged in an attempt to provide an autonomous complete description of the sea---yielding topography and surface conditions.
Further analysis of existing datasets (which could be organized in a coordinated database for the benefit of the community) and future experiments at higher altitudes will continue to refine our understanding of the  potential of this technique.

\setcounter{equation}{0}

% If you have a multiline equation that needs only
% one equation number, use a \nonumber command in
% front of the double backslashes (\\) as shown in
% equations above and below.
%
% ADDITIONAL EQUATION FEATURES
%
% To add letters after equation numbers, place your
% equation or eqnarray within a \begin{mathletters}
% and \end{mathletters} environment.  This environment
% can enclose several equations:

% \begin{mathletters}
% \begin{eqnarray}
% \gamma^\mu & = & \left(\begin{array}{cc} 0 &
% \sigma^\mu_+ \\
% \sigma^\mu_- & 0 \end{array} \right),
% \;\gamma = \left( \begin{array}{cc}
% \! \! -1 & \! \! 0 \\ \! \! 0 & \! \! 1
% \end{array} \right), \\ & & \nonumber \\
% \sigma^\mu_{\pm} & = & ({\bf 1} ,\pm \sigma),
% \end{eqnarray}
% giving
% \begin{eqnarray}
% \not\!a= \left(\begin{array}{cc}0 &
% (\not\!a)_+\\(\not\!a)_- &
% 0\end{array}\right),
% \;(\not\!a)_\pm=a_\mu\sigma^\mu_\pm\;,
% \end{eqnarray}
% \end{mathletters}

%% ------------------------------------------------------------------------ %%
%
%  IN-TEXT LISTS
%
%% ------------------------------------------------------------------------ %%

% Do not use bulleted lists; enumerated lists are okay.
% begin{enumerate}
% \item
% \item
% \item
% \end{enumerate}

%% ------------------------------------------------------------------------ %%
%
%  ACKNOWLEDGMENTS
%
%% ------------------------------------------------------------------------ %%

\begin{acknowledgments}
This study was carried out under the ESA contract TRP~ETP~137.A.  We thank  EADS-Astrium and all sub-contractors (CLS, IEEC, IFREMER, GMV) for their collaboration in the project, and the Institut Cartografic de Catalunya for  flawless flight operations and aircraft GPS/INS kinematic processing. The buoy measurement was provided by IEEC. Finally, we thank  CRESTech, for providing us with SSM/I Integrated Water Vapor data. 

{\em All Starlab authors have contributed significantly; the Starlab author list has been ordered randomly.}
\end{acknowledgments}

\end{article}

\end{document}